\documentclass[12pt]{article}
\usepackage{amssymb}
\usepackage{latexsym}
\usepackage{graphicx}

\renewcommand{\thefootnote}{\fnsymbol{footnote}}

\def\appendix#1{
  \addtocounter{section}{1}
  \setcounter{equation}{0}
  \renewcommand{\thesection}{\Alph{section}}
  \section*{Appendix}
  \addcontentsline{toc}{section}{Appendix \thesection\ \ \ #1}
  }

\newcommand{\newsection}{    
\setcounter{equation}{0}
\section}

\def\bea{\begin{eqnarray}}
\def\eea{\end{eqnarray}}

\def\be{\begin{equation}}
\def\ee{\end{equation}}

\def \bi{\bibitem}

\def\e9{\mbox{${E_9}$}}

\def\id{\protect{{1 \kern-.28em {\rm l}}}}

\begin{document}

\setcounter{page}{1}
\renewcommand{\thefootnote}{\arabic{footnote}}
\setcounter{footnote}{0}

\begin{titlepage}
\begin{flushright}
CU-TP-1013\\
\end{flushright}
\vspace{1cm}

\begin{center}
{\LARGE Two Black Hole Holography, Lensing and Intensity}

\vspace{1.1cm}
{\large{Mark G. Jackson}\footnote{
E-mail: markj@phys.columbia.edu} }\\

\vspace{18pt}

{\it Department of Physics}

{\it Columbia University}

{\it  New York City, NY 10027}
\\
\end{center}
\vskip 0.6 cm

\begin{abstract}
We numerically verify the analysis of the ``expanding horizon" theory of Susskind in 
relation to the 't Hooft holographic conjecture.  By using a numerical 
simulation to work out the holographic image formed by two black holes upon a screen 
very far away, it is seen that it is impossible for a horizon to hide 
behind another.  We also compute the holographic intensity distribution of such an arrangement.
\end{abstract}
\end{titlepage}
\newsection{Introduction}
\paragraph{} In 1988 't Hooft made the surprising announcement that 
the world had one less dimension than previously believed \cite{tho}.  This 
was based on the number of degrees of freedom in a given region of space, 
being found not to depend on the volume, as might be naively expected, but 
rather on the surface area containing the volume.  Furthermore, this conclusion 
connected with one of the few nontrivial combinations of gravity and quantum 
field theory, the entropy content of a black hole \cite{bek}.  't Hooft went on to claim that 
this idea of dimensional reduction, or ``holography," would need to be 
present in any unifying theory.
\paragraph{} At the time of 't Hooft's hypothesis string theory was not yet understood at a level that such 
a claim could be tested in that area (indeed, Maldacena later found 
precisely this feature \cite{malda}).  Susskind was able to show that 
this claim had validity in General Relativity \cite{sus};  if it was true that the 
entropy density of a surface was maximal on that bounding a black hole, 
then this area must expand when projected onto a screen far away.  
In this computation we count only the geodesics orthogonal to the screen, 
since we are not allowed any information which may depend on the screen's 
position; a geodesic which is nonperpendicular to the screen will intersect 
in different locations depending on the screen's position.
This ``expanding horizon" theory of Susskind was explicitly verified by Corley and Jacobson \cite{cj} for the 
single black hole case.  But another observation of Susskind's was that 
the information in a second black hole was not allowed to hide behind the 
first, with respect to the screen; the geodesics will arrange themselves to 
project all information to the screen with non-increasing entropy 
density.  Corley and Jacobson quote a magnification factor of the black 
hole image at large separation, but no examples were given.  In this paper 
we explicitly calculate the geodesics in a two Schwarzschild black hole
geometry.  Black hole lensing has been carried out previously 
\cite{other1} \cite{other2} but to our knowledge not for two black holes, 
which has the novel feature of an arbitrary number of orbits between them 
before the photon escapes.  Furthermore, the geometry used easily lends itself to a 
calculation of the holographic intensity pattern formed by a uniformly radiating black 
hole (the intensity pattern formed keeping only the geodesics 
perpendicular to the screen).  We also perform this calculation.
\newsection{Single Black Hole}
\subsection{Calculation of Orbits}
\paragraph{} Beginning with the Schwarzschild black hole metric normalized to $r=1$ at 
the horizon,
\begin{equation}
ds^2 = \left( 1 - \frac{1}{r} \right) dt^2 - \frac{dr^2}{1 - 1/r} - r^2 d 
\theta ^2 - r^2 \sin ^2 \theta d \phi^2,
\end{equation}
the null geodesic equation can be put into the form $(s \equiv 1/r)$
\begin{equation}
d \phi = \frac{ds}{ \sqrt{b^{-2} - s^2 (1-s) } } 
\end{equation}
where $b$ is the impact parameter and $\phi$ is the azimuthal angle in the
plane of our choosing.  It is most useful to do the 
integration ``backward," in the sense that we begin at the screen and track 
the geodesic until it has hit the horizon or has deflected away from the 
black hole.  We will first work out the case where the geodesic hits the 
horizon, then calculate the case where it deflects off one black hole to 
hit another.
\begin{figure}[t]
\begin{center}
\includegraphics[scale=0.5]{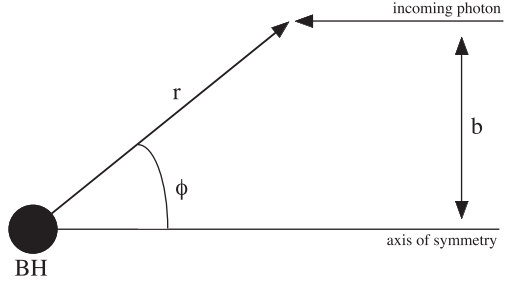}
\caption{A photon approaching a single Schwarzschild black 
hole from the screen infinitely far away.}
\end{center}
\end{figure}
\paragraph{}
Integrating from $r=\infty$ to $r = 1$ means $s$ goes from 0 to 1.  Using 
the photon's incoming line of flight to define $\phi = 0$, we get
\begin{equation}
\phi (b) = \int ^1 _0 \frac{ds}{ \sqrt{b^{-2} - s^2 (1-s) } }.
\end{equation}
\paragraph{}
The denominator vanishes at $b_c=\frac{3 \sqrt{3}}{2}$, indicating that is the 
largest value of $b$ for which we can establish a map between the screen 
and the black hole.  Before that, though, the integral (or angle of the 
photon upon absorption) goes from 0 to infinity,  implying that there 
exists some $b$ for which the photon will orbit around the black hole 
indefinitely before finally being absorbed.  Corley and Jacobson \cite{cj} define a 
``covering" to be one complete image of the horizon upon the screen; the 
first covering is due to emission from the black hole between $\phi = 0$ and 
$\phi = \pi/2$, the second covering from $\phi=\pi/2$ to $\phi = \pi$, as 
shown in figure 2.
\begin{figure}[t]
\begin{center}
\includegraphics[scale=0.3]{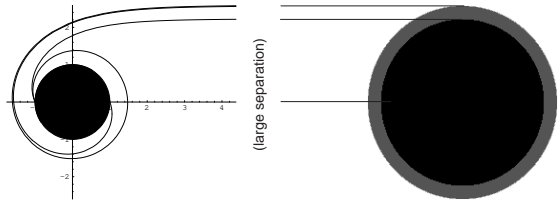}
\caption{Null geodesics showing boundaries of the first few 
coverings (left) and the corresponding screen image (right).}
\end{center}
\end{figure}
\paragraph{} We obtained the result found in \cite{cj} that the maximal impact parameter 
for the first covering is $b \approx 2.23$; this represents an area of 15.60, larger 
than the horizon area $4 \pi \approx 12.57$.  Successive coverings are 
smaller than this on the screen.
\subsection{Intensity}
\paragraph{} We now compute not just the location of the coverings on
the screen, but the intensity pattern a ``holographic viewer" would 
observe, i.e. a viewer only capable 
of observing geodesics perpendicular to the screen.  This is based upon a black 
hole radiating uniformly on its surface, then calculating the density of 
geodesics orthogonal to the screen.  The first step is to calculate the 
angle $\psi$ at which the geodesic hits the black hole,
\begin{equation}
\left. \tan \psi \right|_{\it surface}= \left. \frac{r}{dr/d \phi} 
\right|_{s=1}.
\end{equation}
For us,
\begin{eqnarray}
\nonumber
\frac{dr}{d \phi} &=& \frac{dr}{ds} \frac{ds}{d \phi}, \\
&=& \frac{1}{s^2} \sqrt{b^{-2} - s^2 (1-s) }.
\end{eqnarray}
\paragraph{} Thus we obtain $\left. \tan \psi \right|_{s=1} = b$. 
\begin{figure}[t]
\begin{center}
\includegraphics[scale=0.2]{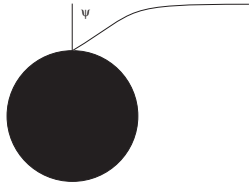}
\caption{A geodesic will hit the black hole at an angle.}
\end{center}
\end{figure}
\paragraph{}  The intensity of a point on the screen will be the ratio of 
geodesic phase space between the screen and the corresponding point on the 
black hole, so we must allow our photon to have small variations from 
the geodesic path.  We begin with the standard case of a ray leaving 
$\psi(b)$ from the normal (see Figure 3) and leaving the BH in the 
plane at angle $\phi(b)$; let us now think of $\psi$ as the independent 
variable and refer to $\phi(b)$ as $\phi(\psi)$.  The geodesic impacts the 
screen at $(x,y,z) = (0, b(\psi), Z) = (0, \tan \psi, Z)$.  Now rotate the 
BH clockwise in the z-y plane by angle $\phi (\psi)$, making the ray hit the screen at
\begin{equation}
\left(
\begin{tabular}{c}
$x $ \\
$y $ \\
$z $ 
\end{tabular}
\right) =
\left(
\begin{tabular}{ccc}
$1$ & $0$ & $0$ \\
$0$ & $\cos \phi (\psi) $ & $- \sin \phi (\psi)$  \\
$0$ & $\sin \phi (\psi)$ & $\cos \phi (\psi)$
\end{tabular}
\right)
\left(
\begin{tabular}{c}
$0$ \\
$\tan \psi$ \\
$Z$ 
\end{tabular}
\right) 
\end{equation}
so it appears as though the ray were leaving from the BH on the $z$ axis;  we will 
call this transformation $J_1$.  Now rotate 
around the z axis so that the ray is leaving at an angle $\theta _2$ from 
the y-z plane,
\begin{equation}
J_2 = \left(
\begin{tabular}{ccc}
$\cos \theta _2 $ & $- \sin \theta _2 $ & $0$ \\
$\sin \theta _2 $ & $\cos \theta _2 $ & $0$ \\
$0$ & $0$ & $1$ 
\end{tabular}
\right).
\end{equation}
Rotate back by arbitrary $\phi$ in the y-z plane
\begin{equation}
J_3 = \left(
\begin{tabular}{ccc}
$1$ & $0$ & $0$ \\
$0$ & $\cos \phi $ & $\sin \phi$  \\
$0$ & $- \sin \phi$ & $\cos \phi$  
\end{tabular}
\right).
\end{equation}
Finally rotate around the z-axis again by arbitrary angle $\theta$,
\begin{equation}
J_4 = \left(
\begin{tabular}{ccc}
$\cos \theta $ & $- \sin \theta $ & $0$ \\
$\sin \theta $ & $\cos \theta$ & $0$ \\
$0$ & $0$ & $1 $ 
\end{tabular}
\right).
\end{equation}
Then the final location of the photon is
\begin{equation}
\left(
\begin{tabular}{c}
$x $ \\
$y $ \\
$z $ 
\end{tabular}
\right) =
J_4 J_3 J_2 J_1
\left( 
\begin{tabular}{c}
$0$ \\
$\tan \psi$ \\
$Z$ 
\end{tabular}
\right) .
\end{equation}
Thus we can define our new impact parameter $b' \equiv \sqrt{x^2 + 
y^2}$, but clearly upon $\theta = \theta _2 = 0, \phi = \phi(\psi) = \phi (b), \psi = 
\psi (b)$ where $b$ is some parameter of our choosing, we recover $b' = b$.  We also take the $Z \rightarrow \infty$ 
limit since that is where the screen is.  To compare 
the areas of the BH and screen, we compute
\begin{eqnarray}
\nonumber
\left. \frac{d (\cos \phi) \ d \theta}{b' \ db' \ d\theta} \right| _{\theta = \varphi = 0, \phi = \phi (b), \psi = 
\psi (b'), b'=b,Z \rightarrow \infty} &=& \left. \frac{\sin \phi}{b'} \left( \frac{db'}{d \phi} \right) ^{-1} \right| _{...} \\
\nonumber &=& \left. \frac{\sin \phi}{b'} \left( \frac{1}{\sqrt{x^2 + y^2 }} 
\left( x \frac{dx}{d \phi} + y \frac{dy}{d \phi} \right) \right) ^{-1} \right| 
_{...} \\
\nonumber&=& \left. \frac{\sin \phi (b)}{b} \left( Z + O(1/Z^2) \right) ^{-1} \right| _{z \rightarrow \infty} \\
&=& \frac{1}{Z} \frac{\sin \phi (b)}{b} .
\end{eqnarray}
We now examine at how the photon's tangent vector at the screen changes under these 
transformations.  We know it begins hitting the screen perpendicular, and 
then will undergo the same transformations as the position vector:
\begin{equation}
\left(
\begin{tabular}{c}
$x $ \\
$y $ \\
$z $ 
\end{tabular}
\right) =
J_4 J_3 J_2 J_1
\left( 
\begin{tabular}{c}
$0$ \\
$0$ \\
$1$ 
\end{tabular}
\right) .
\end{equation}
Now define the angle that it hits the screen at as $\sin \psi'  \equiv 
\sqrt{x^2 + y^2}$ and the angle out of the y-z plane as $\tan \varphi  
\equiv x/y$.  Then we have the following tangent vector phase space comparison between 
the ray leaving the black hole and that hitting the screen:
\begin{eqnarray}
\nonumber \left. \frac{d (\cos \psi) \ d \theta _2}{d (\cos \psi ') \ d \varphi } 
\right|_{...} &=& \left. \frac{ \sin \psi}{ 
\sin \psi ' } \left( \frac{d \psi ' }{d \psi} \right)^{-1} \left( \frac{d 
\varphi}{d \theta _2} \right)^{-1} \right| _{...}, \\
\nonumber &=& \frac{\sin \psi}{\sin \psi ' } \left(  \frac{1}{\sqrt{1 - x^2 - y^2 }} \frac{1}{\sqrt{x^2 + y^2 }} 
\left( x \frac{dx}{d \psi} + y \frac{dy}{d \psi} \right) 
\right)^{-1} ,\\
\nonumber &\times& 
\left( \frac{1}{1 + (x/y)^2} \left( \frac{1}{y} \frac{dx}{d \theta _2} - 
\frac{x}{y^2} \frac{dy}{d \theta _2} \right) \right)^{-1} ,\\
\nonumber &=& \frac{b}{\sqrt{1 + b^2} \sin \phi (b) \frac{d \phi (b)}{d \psi} } ,\\
\nonumber &=& \frac{b}{\sqrt{1 + b^2} \sin \phi (b) \frac{d \phi (b)}{db} 
\frac{db}{d \psi} } , \\
&=& \frac{b}{ (1 + b^2)^{3/2} \sin \phi (b) \frac{d \phi (b) }{db} }.
\end{eqnarray}
\begin{figure}[t]
\begin{center}
\includegraphics[scale=0.5]{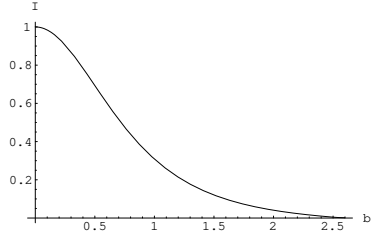}
\caption{Normalized intensity as a function of impact parameter in a 
single black hole.}
\end{center}
\end{figure}
\begin{figure}[t]
\begin{center}
\includegraphics[scale=0.5]{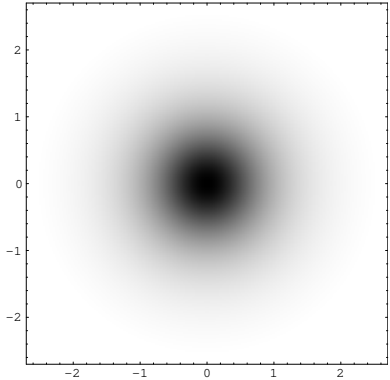}
\caption{The intensity pattern formed on the screen due to a 
single black hole.}
\end{center}
\end{figure}
Combining the phase-space ratios we get the intensity:
\begin{eqnarray}
\nonumber I(b) &=& \frac{\textrm{total phase space of rays leaving 
BH}}{\textrm{total phase space 
of rays hitting screen}} , \\
\nonumber &=& \frac{d (\cos \phi) \ d \theta \ d (\cos \psi) \ d \theta _2}{b' \ db' \ 
d\theta \ d (\cos \psi ') \ d \varphi} , \\
&=& \frac{1}{Z} \frac{1}{ (1 + b^2)^{3/2} \frac{d \phi (b) }{db} }.
\end{eqnarray}
The result is shown in Figures 4 and 5.  In computing the image we have factored 
out the $1/Z$ since this represents the obvious 
dependence on distance and we only care about the intensity normalized at 
some specific distance. 
\newsection{Two Black Holes}
\subsection{Calculation of Orbits}
\paragraph{} We now use the techniques developed in the previous section 
to construct the metric for two black holes; this metric is valid for 
large separation between the black holes.  The axes are chosen so 
the screen lies in the $x-y$ plane, and the origin is at the position of the 
first black hole (the one nearer the screen) with the screen is a very large 
distance down the positive $z$ axis.  The second black hole is taken to 
lie at the point
\begin{equation}
BH_2 = (\rho \cos \mu, \rho \sin \mu, -D).
\end{equation}
We construct the scattering process in reverse by considering photons 
emitted in the normal $(-{\hat {\bf z}})$ direction from the screen, track 
their geodesics, and pick out those which are absorbed by one of the two 
black holes.  Since varying $\mu$ above rotates the picture about the $z$ 
axis we consider only photons emitted from points on the screen with 
$y=0$.  So the trajectory describing the scattering off the first black 
hole lies entirely in the $x-z$ plane.
\begin{figure}[t]
\begin{center}
\includegraphics[scale=0.4]{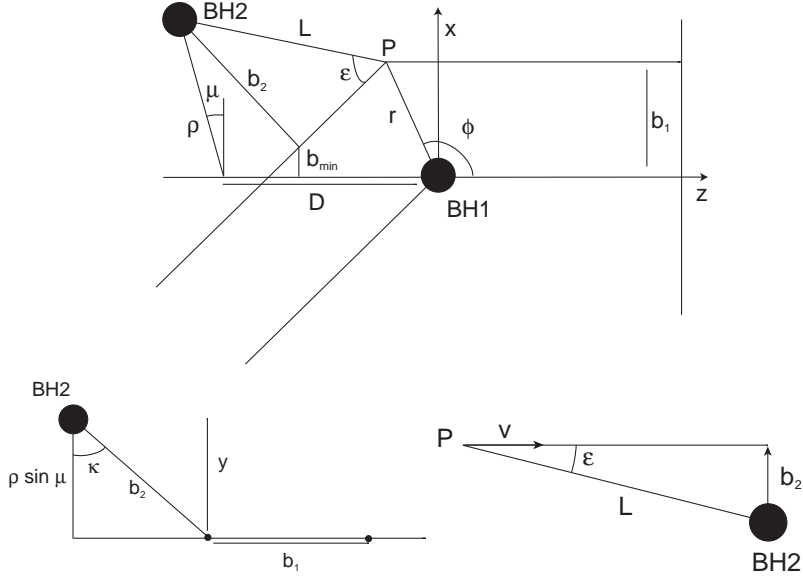}
\caption{Obtaining the impact parameter for the second black hole.}
\end{center}
\end{figure}
\paragraph{} The photon is emitted at a distance $b_1$ from the $z$ axis.  It is 
deflected with angle $\phi (b)$ by the first black hole.  The point $P$ 
at which the asymptotes meet is
\begin{equation}
P = (r \sin \phi, 0, r \cos \phi)
\end{equation}
where
\begin{equation}
r = \frac{b}{\sin \phi }
\end{equation}
as evident from the triangle indicated.  The direction of the deflected 
photon (the lower asymptote in the figure) is given by the unit vector
\begin{equation}
{\bf v} = (\sin 2 \phi , 0, \cos 2 \phi ).
\end{equation}
Our task is now simply to compute the distance $b_2$ of the line with these 
direction cosines passing through the point $P$ from the point $BH_2$.  
The photon will be absorbed by the second black hole if $b_2 < b_c$.  
\paragraph{} Drawing the line connecting $P$ and $BH_2$, the total length 
is $L$ and there is an angle $\epsilon$ with ${\bf v}$, making the answer
\begin{equation}
b_2 = L \sin \epsilon .
\end{equation}
The vector ${\bf L}$ joining them is
\begin{equation}
{\bf L} = BH_2 - P = (\rho \cos \mu - r \sin \phi , \rho \sin \mu, -D - r 
\cos \phi)
\end{equation}
so its length satisfies
\begin{equation}
L^2 = D^2 + \rho ^2 + r^2 - 2 \rho r \cos \mu \sin \phi + 2 D r \cos \phi.
\end{equation}
To determine $\epsilon$ we use
\begin{equation}
\cos \epsilon = \frac{ {\bf L} \cdot {\bf v} }{L}.
\end{equation}
The inner product is also simple to compute:
\begin{eqnarray}
\nonumber L \cos \epsilon &=& {\bf L} \cdot {\bf v} =  (\rho \cos \mu - r \sin 
\phi ) \sin 2 \phi - (D + r \cos \phi) \cos 2 \phi \\
\nonumber &=&  \rho \cos \mu \sin 2 \phi - D \cos 2 \phi - r (\sin \phi \sin 2 
\phi + \cos \phi \cos 2 \phi ) \\
&=&  \rho \cos \mu \sin 2 \phi - D \cos 2 \phi - r \cos \phi.
\end{eqnarray}
From these we compute simply
\begin{equation}
b_2 = L \sin \epsilon = \sqrt{L^2 - (L \cos \epsilon)^2 }.
\end{equation}
This impact parameter is then fed into our previous calculations for a 
single black hole, with appropriate distance normalization due to the mass 
of the second black hole; remember in our units 1 = 1 horizon unit, and all 
calculations in this geometry were done based on the horizon of the first 
black hole.  It is immediately obvious that this procedure can be 
generalized to any number of black holes and ``back-and-forth" orbits 
between them.  To get a concrete visualization of this process, we show in 
Figure 7 the case with $D=10$, $0 \leq \rho \leq 20$.  In Figure 8 we compute 
the area of the first covering on the second black hole for $D=10, 0 \leq 
\rho \leq 25$.  For this we included only the first-covering area to the 
left of the first black hole, as there is a redundency on the right side 
and we are only interested in how much area it takes to represent the 
whole horizon once.  The hiding theorem is clearly verified, the horizon 
expanding even more when it forms a ring.
\begin{figure}[t]
\begin{center}
\includegraphics[scale=0.6]{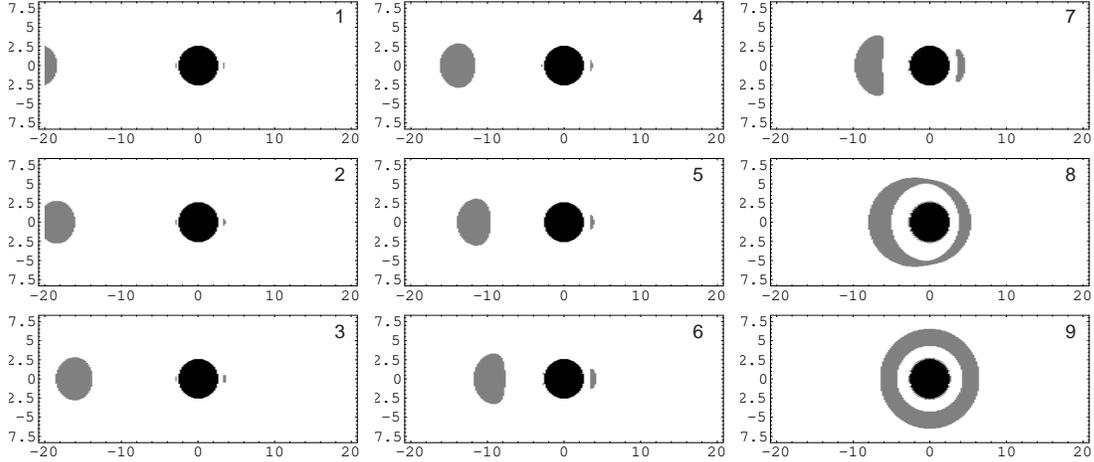}
\caption{Attempted horizon hiding for $D=10,$ $0 \leq \rho \leq 20$.  Gray 
represents the geodesic hitting the second black hole, black indicates hitting the first.}
\end{center}
\end{figure}
\begin{figure}[t]
\begin{center}
\includegraphics[scale=0.5]{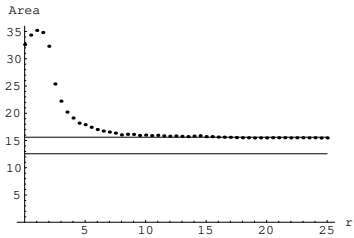}
\caption{The area of the first covering when attempting to 
hide.  The lower bar indicates the horizon area $= 4 \pi$, which we need to 
be above.  The upper bar indicates the area of the first covering for a 
single black hole, which the second black hole must approach at large 
separation.}
\end{center}
\end{figure}
\subsection{Intensity}
\paragraph{} Thinking about phase space as before, we see that for two black holes the 
situation is very similar except we now remember the rays leave the ``real" 
screen, deflect off BH1 to form a ``virtual screen," then hit BH2.  Thus
\begin{eqnarray}
\nonumber I &=& \frac{\textrm{phase space of rays leaving 
BH 2 }}{\textrm{phase space 
of rays hitting real screen}} ,\\
\nonumber &=& \frac{\textrm{PS 
of rays leaving BH 2}}{\textrm{PS 
of rays hitting virtual screen}} 
\frac{\textrm{PS
of rays hitting virtual screen}}{\textrm{PS 
of rays hitting real screen}} ,\\
&=& \frac{1}{ (1 + b_2 ^2)^{3/2} \frac{d \phi (b_2) }{db_2} } 
\frac{\textrm{differential area of virtual screen}}{\textrm{differential area 
of real screen} }.
\end{eqnarray}
where the first part is from our previous single BH analysis, and the 
second is just comparing the differential area containing the ray on the 
virtual screen versus the real screen.  Note that we have omitted the 
phase space on the screens due to the tangent vector in this step, but this 
is just unity because both screens by definition only take only 
perpendicular rays and
so have the same tangent phase space.  The ratios of differential areas are 
then
\begin{equation}
\frac{b_2 \ db_2 \ d \kappa}{b_1 \ db_1 \ d \theta}
\end{equation}
where $\kappa$ is shown in Fig. 6 and $\theta$ is the usual coordinate 
around the z axis.  We evaluate this 
by switching to cartesian coordinates $(b_1,y)$ for
BH2:
\begin{equation}
b_2 \ db_2 \ d \kappa \rightarrow db_1 \ dy
\end{equation}
\begin{figure}[t]
\begin{center}
\includegraphics[scale=0.6]{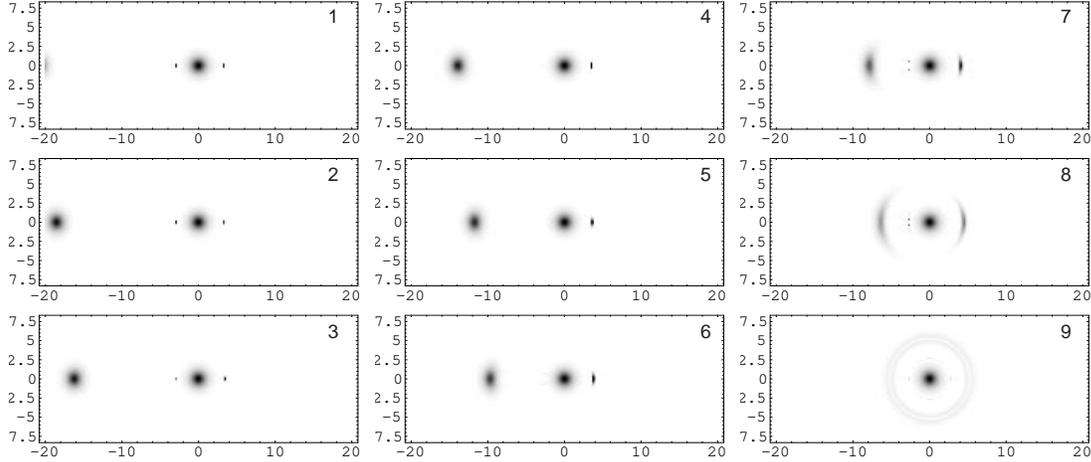}
\caption{As a black hole hides behind another, the image becomes very faint.}
\end{center}
\end{figure}
Now see that this yields immediate cancellation with the $b_1$.  As for $dy$, 
we see that this point being lifted out of the plane is a distance 
$b_{min}$ from the z-axis, thus $dy = b_{min} d \theta$.  The $d \theta$'s 
cancel, leaving only $b_{min} / b_1$.  With
\begin{equation}
b_{min} = b_1 - L \cos \epsilon \sin (2 \phi - \pi)
\end{equation}
and the expression previously obtained for $L \cos \epsilon$ we 
get 
\begin{equation}
\frac{b_{min}}{b_1} = 1 + \frac{L \cos \epsilon \sin 2 \phi}{b_1}.
\end{equation}
So the total intensity is just
\begin{equation}
I = \frac{1}{ (1 + b_2 ^2)^{3/2} \frac{d \phi (b_2) }{db_2} }  \left( 1 + 
\frac{L \cos \epsilon \sin 2 \phi (b_1)}{b_1} \right).
\end{equation}
We now repeat the previous lensing calculation, but including this intensity 
profile.  The result is shown in Figure 9.
\newsection{Conclusion}
\paragraph{} We have verified Susskind's hypothesis that the entropy is 
maximal on the surface of a black hole.  We did this for the two black hole case and found that no 
entropy-information is lost when one black hole attempts to 
hide behind the other.  The holographic intensity of such a configuration was also 
calculated.
\newsection{Acknowledgements}
\paragraph{} The author thanks B. Paczynski, H. Peiris, E. Weinberg, and especially R. Plesser for helpful comments.

\end{document}